\documentclass[twocolumn,showpacs,preprintnumbers,amsmath,amssymb]{revtex4}

\usepackage{graphicx}
\usepackage{dcolumn}
\usepackage{bm}
\usepackage{upgreek}

\begin{document}


\title{Constriction-limited detection efficiency of superconducting nanowire single-photon detectors}

\author{Andrew J. Kerman}
\affiliation{Lincoln Laboratory, Massachusetts Institute of Technology, Lexington, MA, 02420}
\author{Eric A. Dauler, Joel K.W. Yang, Kristine M. Rosfjord, Vikas Anant, and Karl K. Berggren}
\affiliation{Research Laboratory of Electronics, Massachusetts Institute of Technology, Cambridge, MA, 02139}
\author{G. Gol'tsman and B. Voronov}
\affiliation{Moscow State Pedagogical University, Moscow 119345,
Russia}

\date{\today}

\begin{abstract}
We investigate the source of large variations in the observed
detection efficiencies of superconducting nanowire single-photon
detectors between many nominally identical devices. Through both
electrical and optical measurements, we infer that these variations
arise from ``constrictions:" highly localized regions of the
nanowires where the effective cross-sectional area for
superconducting current is reduced. These constrictions limit the
bias current density to well below its critical value over the
remainder of the wire, and thus prevent the detection efficiency
from reaching the high values that occur in these devices only when
they are biased near the critical current density.
\end{abstract}

\maketitle

Superconducting nanowire single-photon detectors (SNSPDs)
\cite{newtech,newsob,inductance,cavity} provide a unique combination
of high infrared detection efficiency (up to 57\% at 1550nm
\cite{cavity} has been demonstrated) and high speed ($<$30 ps timing
resolution \cite{newsob,ASCeric}, and few-ns reset times after a
detection event \cite{inductance}). Applications for these devices
already being pursued include high data-rate interplanetary optical
communications \cite{comm}, spectroscopy of ultrafast quantum
phenomena in biological and solid-state physics
\cite{nistfiber,qspec}, quantum key distribution (QKD) \cite{QKD},
and noninvasive, high-speed digital circuit testing \cite{CMOS}.

In many of these applications, large arrays of SNSPDs would be
extremely important \cite{ASCeric}. For example, existing SNSPDs
have very small active areas, making optical coupling relatively
difficult and inefficient \cite{nistfiber,sobfiber}. Their small
size also limits the number of optical modes they can collect, which
is critical in free-space applications where photons are distributed
over many modes, such as laser communication through the atmosphere
(where turbulence distorts the optical wavefront) and in
fluorescence detection. Furthermore, it was shown in Ref.
\cite{inductance} that the maximum count rate for an individual
SNSPD decreases as its active area is increased, due to its kinetic
inductance, forcing a tradeoff between active area and high count
rates. Count rate limitations are particularly important in optical
communications and QKD, affecting the achievable receiver
sensitivity or data rate \cite{comm,twoelement}. Detector arrays
could provide a solution to these problems, giving larger active
areas while simultaneously \textit{increasing} the maximum count
rate by distributing the flux over many smaller (and therefore
faster) pixels. Large arrays could also provide spatial and
photon-number resolution. Although few-pixel detectors have been
demonstrated \cite{ASCeric,twoelement,QKD,sobfiber}, fabrication and
readout methods scalable to large arrays have not yet been
discussed.

A first step towards producing large arrays of SNSPDs is to
understand (and reduce) the large observed variation of detection
efficiencies for nominally identical devices \cite{cavity}, which
would set a crippling limit on the yield of efficient arrays of any
technologically interesting size. In this Letter, we demonstrate
that these detection efficiency variations can be understood in
terms of what we call ``constrictions:" highly localized,
essentially pointlike regions where the nanowire cross-section is
effectively reduced, and which are \textit{not} due to lithographic
patterning errors (line-edge roughness).

The electrical operation of these detectors has been discussed
previously by several authors
\cite{newtech,newsob,inductance,cavity,joelmodel}, so we only
summarize it here. The NbN nanowires are biased with a DC current
$I_\mathrm{bias}$ slightly below the critical value $I_\mathrm{C}$.
An incident photon of sufficient energy can produce a resistive
``hotspot'' which in turn disrupts the superconductivity across the
wire, producing a series resistance which then expands in size due
to Joule heating \cite{inductance,joelmodel}. The series resistance
quickly becomes $\gg$50$\Omega$, and the current is diverted out of
the device and into the 50$\Omega$ transmission line connected
across it, resulting in a propagating voltage pulse on the line. The
device can then cool back down into the superconducting state, and
the current through it recovers with the time constant
$L_\textrm{k}/50\Omega$, where $L_\textrm{k}$ is the kinetic
inductance \cite{inductance}.

\begin{figure}
\includegraphics[width=3.25in]{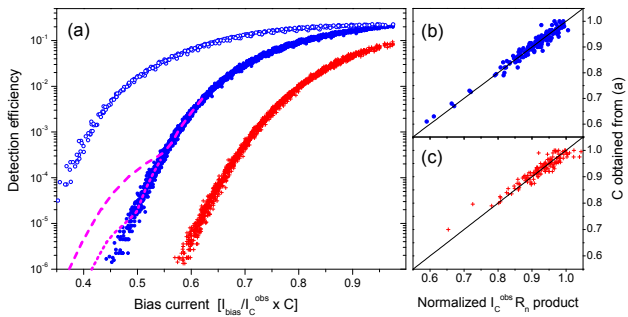}
\caption{Figure \ref{fig:2}: (color online) Constriction values
extracted using DE vs. $I_\textrm{bias}$ data. (a) Universal DE
curves for devices with 90nm wide wires, at $T=1.8$K ($\bullet$) and
$T=4.2$K (+). Data from 170 devices distributed over two separate
fabrication runs is shown superposed. By rescaling the
$I_\textrm{C}^\textrm{obs}$ of each device such that all data lies
on a single universal curve as shown, the constriction $C$ (which
indicates the fraction of the wire's cross-sectional area that is
superconducting at the constricted point) can be obtained. Also
shown ($\circ$) are data for $15$ devices having 54 nm wide wires
(and 36\% fill factor) at $T=1.8$K, indicating that narrower wires
exhibit a very different universal curve shape. These data provide
evidence for the localized nature of the constrictions, since any
appreciably long section of wire having a smaller cross-section
should significantly alter the shape of the curves, making it
impossible to superpose them as shown. This is illustrated by the
broken lines, which show a simple estimate for a 90 nm wide wire
having a 54 nm wide constriction with a length of 0.5 $\upmu$m
(dashed line) or 50 nm (dotted line). These estimates are obtained
simply by adding together the universal curves for the two wire
widths, in a ratio given by the length of wire with each width, i.e.
(0.5 $\upmu$m/50 $\upmu$m) and (0.05 $\upmu$m/50 $\upmu$m). Note
that these broken lines terminate at
$I_\textrm{bias}/I_\textrm{C}^\textrm{obs}\times C=0.6=54/90$ due to
the assumed 54 nm constriction of the 90 nm-wide wire. (b) and (c)
Constriction values $C$ obtained from the data in (a) (for (b)
$T=1.8$K and (c) $T=4.2$K) vs. those obtained using the
$I_\textrm{C}^\textrm{obs}R_\textrm{n}$ product. The $C$ values in
both cases are normalized absolutely using
$L_\textrm{k}(I_\textrm{bias})$, as described below (see Fig.
\ref{fig:3}). The solid lines are straight lines through the origin
with slope 1; no fitting was used.\label{fig:2}}
\end{figure}

The nanowires used in this work were patterned at the shared
scanning-electron-beam-lithography facility in the MIT Research
Laboratory of Electronics using a process described in Refs.
\cite{joelfab,inductance,cavity}, on ultrathin ($\sim5$ nm) NbN
films grown at Moscow State Pedagogical University \cite{films}. The
majority of the wires were on average 90 nm in width, and were
fabricated in a meander pattern with 200 $\upmu$m line pitch (45\%
fill factor), subtending an active area of 3 $\times$ 3.3 $\upmu$m
or 10 $\times$ 10 $\upmu$m. Some devices with 54 nm average width
and 150 nm pitch (36\% fill factor) were also fabricated - see Fig.
\ref{fig:2}(a). The devices had critical temperatures
$T_\mathrm{C}\sim$ $9-10$ K, and critical current densities
$J_\mathrm{C}\sim5\times 10^{10}$ A/m$^2$ at $T=1.8$K.

The experiments were performed at MIT Lincoln Laboratory, using the
procedures and apparatus discussed in detail in Refs.
\cite{inductance, cavity}. Briefly, the devices were cooled to as
low as 1.8 K inside a cryogenic probing station. Electrical contact
was established using a cooled 50 $\Omega$ microwave probe attached
to a micromanipulator, and connected via coaxial cable to the
room-temperature electronics. We counted electrical pulses from the
detectors using low-noise amplifiers and a gated pulse counter. To
optically probe the devices, we used a 1550 nm modelocked fiber
laser (with a 10 MHz pulse repetition rate and $\le$1 ps pulse
duration) that was attenuated and sent into the probing station via
an optical fiber. The devices were illuminated from the back
(through the sapphire substrate) using a lens attached to the end of
the fiber which was mounted to an automated micromanipulator. The
focal spot had a measured e$^{-2}$ radius of $\sim$ 25 $\upmu$m.

\begin{figure}
\includegraphics[width=3.25in]{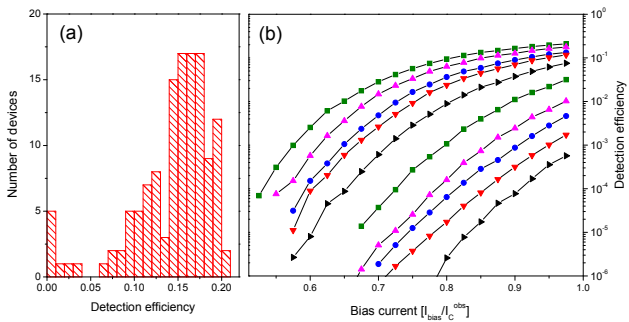}
\caption{Figure \ref{fig:1}: (color online) Variations in SNSPD
detection efficiency. (a) Histogram of the DEs measured for 132
devices from a single fabrication run, on a single chip. The devices
were $3\times 3.3\;\upmu$m in size, and composed of a 50 $\upmu$m
long nanowire in a meander pattern with 45\% fill factor. The
measurements were made at $T=1.8$K, and
$I_\textrm{bias}=0.975I_\textrm{C}^\textrm{obs}$. Note that the peak
DE of 22\% increases to 57\% with the addition of an optical cavity,
as described in Ref. \cite{cavity} (b) Measurements of the DE vs.
$I_\textrm{bias}/I_\textrm{C}^\textrm{obs}$ for a selection of these
devices.\label{fig:1}}
\end{figure}

Figure \ref{fig:1} illustrates the detection efficiency (DE)
variations observed on a single chip of 132 devices of the same
geometry, fabricated in a single run. These devices are the same
ones reported in Ref. \cite{cavity} before the optical cavities were
added (maximum DE after the addition of cavities was 57\%). In panel
(a) we show a histogram of the measured detection efficiencies at
$I_\textrm{bias}=0.975I_\textrm{C}^\textrm{obs}$ (where
$I_\textrm{C}^\textrm{obs}$ is the observed critical current of each
device), and (b) shows some representative data of the observed DE
as a function of $I_\textrm{bias}/I_\textrm{C}^\textrm{obs}$. Note
that the shape of these curves also varies significantly. As we show
below, these data can be explained with the hypothesis, first
suggested in Ref. \cite{constrict}, that some devices have
``constrictions:" regions where the (superconducting)
cross-sectional area $A_\textrm{cs}$ of the wire is reduced by a
factor we label $C$. This effectively reduces the observed critical
current by that same factor
($I_\textrm{C}^\textrm{obs}=J_\textrm{C}A_\textrm{cs}C=I_\textrm{C}C$),
and prevents the current density everywhere but near the
constriction from ever approaching the critical value (and hence
prevents the wire from having a high DE except locally near the
constriction).

If all the nanowires were identical in all dimensions save for a
pointlike constriction, we would expect that if the data of Fig.
\ref{fig:1}(b) were plotted vs. absolute current $I_\textrm{bias}$
(rather than $I_\textrm{bias}/I_\textrm{C}^\textrm{obs}$) it would
all lie on a single, universal curve, with the data for more
constricted devices simply not extending to as high currents. This
turns out to be approximately true, but variations from device to
device across the chip either in film thickness or in the nanowire
width obscure this essential feature of the data when it is plotted
in this simple way. Instead, we present our results as shown in Fig.
\ref{fig:2}. In panel (a), the DE data for each of 170 devices with
90 nm wide wires (across two chips fabricated in separate runs) is
shown superposed (filled circles indicate $T=1.8$K, and crosses
$T=4.2$K). All of the data for each temperature can be made to lie
on a single universal curve by scaling the critical current
$I_\textrm{C}^\textrm{obs}$ for each device by an adjustable factor
(which is just $1/C$: $I_\textrm{C}=I_\textrm{C}^\textrm{obs}/C$).
The very fact that data from this many devices can be so well
superposed is already suggestive of a universal shape. However, we
can now take the $C$ value for each device extracted using this
scaling procedure and cross-check it. Based on our previous
discussion, if all wires were identical save for constrictions, we
would expect $C=I_\textrm{C}^\textrm{obs}/I_\textrm{C}$, i.e. $C$ to
be exactly proportional to $I_\textrm{C}^\textrm{obs}$. Due to the
variations across a chip discussed above, this is only partially
true. However, we can normalize out these variations using a very
simple method: instead of comparing $C$ directly to
$I_\textrm{C}^\textrm{obs}$, we instead compare it to the product
$I_\textrm{C}^\textrm{obs}R_\textrm{n}= (C\times
J_\textrm{C}A_\textrm{cs})(\rho_\mathrm{n}l/A_\textrm{cs})=C\times
J_\textrm{C}\rho_\mathrm{n}l$, where $A_\textrm{cs}$ and
$R_\textrm{n}$ are the cross-sectional area and room-temperature
resistance of each nanowire, $J_\textrm{C}$ and $\rho_\mathrm{n}$
are the critical current density and room-temperature resistivity of
NbN, respectively, and $l$ is the total wire length. This product
depends on the wire geometry only through $l$ (which is fixed
lithographically and does not vary appreciably between devices) and
not on each wire's individual $A_\textrm{cs}$. Figures
\ref{fig:2}(b) and (c) show a comparison between the $C$ values
extracted from the data in Fig. \ref{fig:2}(a) and the
$I_\textrm{C}^\textrm{obs}R_\textrm{n}$ product. The data lie on a
straight line through the origin, indicating that these two
independent measures of $C$ are mutually self-consistent.

Our data can also be used to infer that the constrictions are
essentially pointlike (i.e. very short in length). The open circles
in Fig. \ref{fig:2}(a) are data for 15 devices with 54 nm wide
wires, and clearly show a dramatically different shape (i.e. high DE
persists to much lower currents than for the wider wires). The
broken lines are estimates, based on the data for 90 nm and 54 nm
wide wires, of what one would expect for a device having 90 nm wide
wire, except at a single constriction 54 nm wide (corresponding to
$C\sim 0.6$) with a length of either 0.5 $\upmu$m (dashed line) or
50 nm (dotted line) long. These curves have a different shape from
the data for 90 nm wide wires, because at low currents the DE is
dominated by the region of wire near the constriction, while at
higher current it becomes dominated by the contribution from the
rest of the wire length. This very different shape should be
distinguishable if it were present, and should prevent the data from
being superposed onto a single curve. The absence of this in our
data at any level above the noise indicates that the constricted
regions are likely much shorter than $\sim$0.5$\upmu$m.

So far, we have in fact only measured constriction in a relative
sense; that is, we have no way to tell if our best devices in fact
have $C=1$. To address this, we can exploit the known dependence of
the kinetic inductance on bias current. Kinetic inductance arises
from energy stored in the effectively ballistic motion of Cooper
pairs; as the current density is increased towards the critical
value, the density of Cooper pairs is depleted, forcing the
remaining pairs to speed up (and therefore store more kinetic energy
per unit volume) to maintain the current. Hence, the kinetic
inductance increases as $J\rightarrow J_\textrm{C}$ \cite{orlando}.
Since the kinetic inductivity locally increases with
$J/J_\textrm{C}$, the total kinetic inductance of the wire provides
a way to determine if the current density is indeed near the
critical value over the whole wire or only at one localized place.
We measured the inductance of our nanowires using a network
analyzer, by observing the phase of a reflected microwave signal as
a function of frequency. We then fit this data using a suitable
electrical model to extract the inductance value. A bias tee was
inserted into the signal path to superpose the desired
$I_\textrm{bias}$ with the network analyzer output. The phase
contributions from the coaxial cable, bias tee, and microwave probe
were removed by probing an \textit{in situ} microwave calibration
standard (GGB Industries CS-5). The microwave power used in this
measurement corresponds to a peak current amplitude of $\le 0.5
\upmu A$, and the critical current measured in the presence of the
microwaves was within 10\% of that measured in their absence ($\sim
20\upmu$A for typical devices at $T=1.8$K).

\begin{figure}
\includegraphics[width=3.25in]{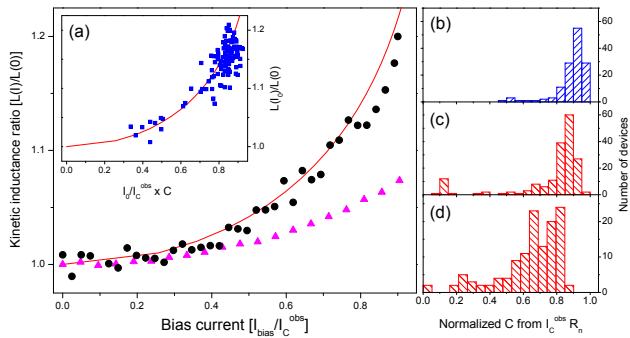}
\caption{Figure \ref{fig:3}: (color online) Absolute measurement of
constriction using bias current-dependence of kinetic inductance.
(a) The kinetic inductance of our nanowires should increase with
$I_\textrm{bias}$, due to the depletion of the Cooper-pair density
near the critical current density (solid line). A detector with the
highest observed detection efficiency on the chip (22\%) behaves as
expected ($\bullet$), with no free parameters. However, a detector
with much lower DE (0.1\%) does not (filled triangles). This is due
to a constriction, which prevents the current density from
approaching critical anywhere but near this one localized point.
Inset: the inductance ratio $\mathcal{R}_\textrm{L}\equiv
L(I_0\approx 0.9I_\textrm{C}^\textrm{obs})/L(0)$ measured for each
device (filled squares), plotted vs.
$I_0/(I_\textrm{C}^\textrm{obs}/C)=I_0/I_\textrm{C}$, where $C$ is
obtained from the $I_\textrm{C}^\textrm{obs}R_\textrm{n}$ product.
These data agree with the prediction (solid line), indicating that
$\mathcal{R}_\textrm{L}$ and $I_\textrm{C}^\textrm{obs}R_\textrm{n}$
give mutually consistent results for the constriction. (b), (c), and
(d): Distribution of $C$ values obtained using purely electrical
measurements, for (b) the same set of devices shown in Fig.
\ref{fig:1}, and (c),(d) for 310 additional devices on a separate
chip, with active areas of both (c) 3 $\times$ 3.3 $\upmu$m and (d)
10 $\times$ 10 $\upmu$m. These data were obtained from
$I_\textrm{C}^\textrm{obs}R_\textrm{n}$, with their normalization
set by single measurements of $L_\textrm{k}(I_\textrm{bias})$ like
that shown in (a), one for each of the three sets of devices shown
in (b), (c), and (d). The devices from (b) and (c) are nominally
identical in all respects, though they were fabricated on different
NbN films, and at different times. Their $C$ distributions are are
therefore quite similar. The difference between (c) and (d)
indicates that larger devices have a significantly higher
probability of a sizeable constriction.\label{fig:3}}
\end{figure}

In Fig. \ref{fig:3}(a), we show the measured inductance vs. current
of two devices; one which has nearly the highest detection
efficiency observed on this chip (22\% - filled circles), and the
other having one of the lowest (0.1\% - filled triangles). Also
shown is the prediction for $L_\textrm{k}(I_\textrm{bias})$ from
Ginsburg-Landau theory (see, e.g., \cite{orlando}), with no free
parameters (solid line). The data for the high-DE device show good
agreement with this prediction, indicating that this device is
indeed unconstricted. However, for the low-DE device the inductance
does not increase as much as predicted, which is precisely what we
would expect within the constriction hypothesis; the current density
is only near critical at one localized place (which constitutes a
negligible fraction of the total wire length) whereas everywhere
else the current density is lower, producing a smaller increase in
inductance. The factor by which $I_\textrm{C}^\textrm{obs}$ must be
rescaled for a given device so that the
$L_\textrm{k}(I_\textrm{bias})$ data matches the prediction
constitutes an absolute measurement of $C$. One such measurement for
a single representative device, from among a large set of nominally
identical devices, then allows us to correctly normalize the $C$
values obtained using either of the previous methods described above
(see Fig. \ref{fig:2}) for all other devices in that set.

We can also verify that the observed $L_\textrm{k}(I_\textrm{bias})$
and $I_\textrm{C}^\textrm{obs}R_\textrm{n}$ product give mutually
consistent results. To check this, for each device we measured the
inductance ratio $\mathcal{R}_\textrm{L}\equiv L(I_0)/L(0)$, where
$I_0\approx 0.9I_\textrm{C}^\textrm{obs}$. Using the $C$ obtained
from the normalized $I_\textrm{C}^\textrm{obs}R_\textrm{n}$ product,
we also obtain $I_0/I_\textrm{C}=I_0C/I_\textrm{C}^\textrm{obs}$ for
each device. The inset to Fig. \ref{fig:3}(a) shows
$\mathcal{R}_\textrm{L}$ vs. $I_0/I_\textrm{C}$ (filled squares),
and the data are in reasonable agreement with the Ginsburg-Landau
prediction (solid line).

In addition to providing evidence for the constriction hypothesis,
the measurement of $L_\textrm{k}(I_\textrm{bias})$ and
$I_\textrm{C}^\textrm{obs}R_\textrm{n}$ provide a powerful
diagnostic tool, since they constitute a purely electrical
measurement of $C$, which can then be used to predict the detection
efficiency, as indicated in Figs. \ref{fig:2}(b) and (c). Since
optical testing of large numbers of detectors is significantly more
difficult and time-consuming than electrical testing, this is
potentially an important screening technique. An example of data
obtained with this technique is shown in Figs. \ref{fig:3}(b), (c),
and (d). For three sets of devices, we obtained $C$ values from
$I_\textrm{C}^\textrm{obs}R_\textrm{n}$, normalized using a single
absolute measurement of $C$ for each set via
$L_\textrm{k}(I_\textrm{bias})$ as described above. Panel (b) shows
data for the set of devices from Fig. \ref{fig:1}(a); panels (c) and
(d) show results for a sample of 310 additional devices on a single
additional chip, all having 90 nm wide wires on a 200 nm pitch, but
with active areas of both (c) 3 $\times$ 3.3 $\upmu$m, and (d) 10
$\times$ 10 $\upmu$m. The devices from (b) and (c), which are
nominally identical, exhibit very similar $C$ distributions (though
some yield fluctuation, which we commonly observe, is evident).
Panels (c) and (d), however, clearly show a lower average $C$ for
the larger-area devices, qualitatively consistent with some fixed
area density of constrictions. This implies that at the present
state of film growth and nanowire patterning, the yield for high-DE
single devices or device arrays covering 10 $\times$ 10 $\upmu$m
areas or larger will likely be quite low.

As a final note, we remark on the obvious question of the origin of
these constrictions. The most natural explanation would be
lithographic patterning errors; for example, a small particle or
defect in the resist before exposure could result in a localized
narrow section of wire. However, we have performed extensive
scanning electron microscopy of devices that were measured to be
severely constricted (e.g. $C\sim 0.5$) and no such errors were
observable. This suggests that constrictions in our devices result
either from thickness variations or material defects, which may have
been present in the film before patterning, and may even be due to
defects present in the substrate itself before film growth.

In conclusion, we have verified both optically and electrically that
the large variations in detection efficiency between nominally
identical superconducting nanowire single photon detectors are the
result of localized constrictions which limit the device current.
Further work is ongoing to pin down the exact source of these
constrictions, with the hope of eventually fabricating large arrays
of these detectors.

This work is sponsored by the United States Air Force under Air
Force Contract \#FA8721-05-C-0002.  Opinions, interpretations,
recommendations and conclusions are those of the authors and are not
necessarily endorsed by the United States Government.

\end{document}